# A note on using finite non-abelian $p$-groups in the MOR cryptosystem

Ayan Mahalanobis

Department of Mathematical Sciences, Stevens Institute of Technology, Hoboken, NJ 07030.
Ayan.Mahalanobis@stevens.edu

**Abstract.** The MOR cryptosystem [9] is a natural generalization of the El-Gamal cryptosystem to non-abelian groups. Using a $p$-group, a cryptosystem was built in [4]. It seems resoanable to assume the cryptosystem is as secure as the El-Gamal cryptosystem over finite fields. A natural question arises can one make a better cryptosystem using $p$-groups? In this paper we show that the answer is no.

## 1  Introduction

There are three cryptosystems widely in use today – 1) RSA 2) The El-Gamal or the Diffie-Hellman key exchange protocol over the group of points on a Elliptic curve over a finite field; also known as the Elliptic curve cryptosystems and 3) The El-Gamal or the Diffie-Hellman key exchange protocol over the multiplicative group of a finite field. The security type of the RSA and the El-Gamal or the Diffie-Hellman over finite fields is the same. The best known attack against them, the index calculus methods, is sub-exponential in nature. On the other hand, the best known attack against the Elliptic curve cryptosystems is exponential in nature. This is the primary difference of Elliptic curve cryptosystems from others. This is a very loose but effective distinction between cryptosystems.

It is natural to ask the question, can we build some other cryptosystem for which the security is exponential too, as in the Elliptic curve cryptosystems? The quest for an answer to the above question led us to look at MOR [9] and $p$-groups as in [4]. In [4] we showed that it is possible to build a cryptosystem which is as secure as the El-Gamal over finite fields, i.e., is of sub-exponential type. In this paper we show that is the best possible scenario when we are working with MOR over finite $p$-groups.

## 2 The MOR cryptosystem

Let $G$ be a group and $\phi : G \to G$ be an automorphism. In this paper, if we work with automorphisms of $G$, we work in the automorphism group of $G$, with the group operation being the composition of automorphisms as mappings.

### 2.1 Description of the MOR cryptosystem

Alice's keys are as follows:

**Public Key** $\phi$ and $\phi^m$, $m \in \mathbb{N}$.
**Private Key** $m$.

*Encryption*

**a** To send a message $a \in G$ Bob computes $\phi^r$ and $\phi^{mr}$ for a random $r \in \mathbb{N}$.
**b** The ciphertext is $(\phi^r, \phi^{mr}(a))$.

*Decryption*

**a** Alice knows $m$, so if she receives the ciphertext $(\phi^r, \phi^{mr}(a))$, she computes $\phi^{mr}$ from $\phi^r$ and then $\phi^{-mr}$ and then from $\phi^{mr}(a)$ computes $a$.

Alice can compute $\phi^{-mr}$ two ways; if she has the information necessary to find out the order of the automorphism $\phi$ then she can use the identity $\phi^{t-1} = \phi^{-1}$ whenever $\phi^t = 1$. Also, she can find out the order of some subgroup in which $\phi$ belongs and use the same identity. However the smaller the subgroup, more efficient the decryption algorithm.

## 3 The MOR cryptosystem using linear operators of vector spaces over finite fields

Let us consider the MOR cryptosystem on a finite dimensional vector space $V$ over a finite field $\mathbb{F}_q$, where $q = p^k$; $p$ is a prime and $k$ a positive integer. This means that in place of $G$ in Section 2.1; we are using a vector space $V$ and in place of the automorphism $\phi : G \to G$, we are using a linear transformation $\phi : V \to V$. We want to understand the security of this MOR, this is not only of academic interest [6] but also of future need in this paper.

It is well known [10, Chapter 2] that if we fix a basis for a $n$-dimensional vector space $V$ over $\mathbb{F}_q$, then corresponding to a linear operator $T : V \to V$ there is a $n \times n$ matrix $A$ over $\mathbb{F}_q$ such that $T(x) = Ax$ for all $x \in V$. Furthermore $T^n(x) = A^n x$, i.e., exponentiation of the map $T$ is the same as the exponentiation of the matrix $A$. Furthermore, notice that in MOR the homomorphisms assumed are automorphisms, so the matrix $A$ is invertible. Hence we claim that the discrete logarithm problem in the group of linear operators over a vector space $V$ is the same as the discrete logarithm problem in $\mathrm{GL}(n, q)$, the group of $n \times n$ non-singular matrices over $\mathbb{F}_q$. We simultaneously claim that this provides sufficient evidence to claim that the MOR cryptosystem on the group of linear operators is the same and has the same security as a El-Gamal cryptosystem over $\mathrm{GL}(n, q)$. The next is to show that the discrete logarithm problem over $\mathrm{GL}(n, q)$ is only as secure as the discrete logarithm problem over some finite extension of $\mathbb{F}_q$. This was actually proved in [6]. So, we claim that if we are using the MOR cryptosystem on linear operators over vector spaces then there is no advantage to be gained in terms of security and since matrix multiplication is harder to compute than that of integer multiplication, we see that there are no practical advantages as well.

## 4 The MOR cryptosystem using *p*-groups

The non-abelian $p$-groups are a very nice candidate for the MOR cryptosystem; because computation is easy with them and there are a lot of $p$-groups, see [12]. A lot is known about the $p$-groups and it seems that one might be able to use them to build a good and secure cryptosystem using MOR. This was the motivation behind [4], to consider the group of unitriangular matrices over finite fields for MOR. The MOR cryptosystem over the group of unitriangular matrices was successful in the sense that it is as secure as the El-Gamal cryptosystem over finite fields; it would be nicer if a "more secure", i.e., an exponential type cryptosystem could be built using other $p$-groups. This means that we are trying to build cryptosystems in which the best known attack has exponential time computational complexity, *viz.* the elliptic curve cryptosystem; not RSA or the cryptosystems using the discrete logarithm problem in finite fields; where sub-exponential time attacks are known.

In this paper we show that it is impossible to do that when we are using finite $p$-groups and the MOR cryptosystem; more specifically we show that if we are working with finite $p$-groups the best attack can always be reduced to an attack on the discrete logarithm problem over finite fields.

For this we define the *Frattini subgroup* of a finite group $G$.

**Definition 1.** *[11, Chapter 5]*
*If $G$ is a group, then its Frattini subgroup $\Phi(G)$ is the defined as the intersection of all maximal subgroups of $G$.*

It is well known that $\Phi(G)$ is a characteristic subgroup and hence a normal subgroup of $G$. We now state without proof the following well known theorem.

**Theorem 1.** *[11, Theorem 5.48]*
*Let $G$ be a finite group.*

*(i) $\Phi(G)$ is nilpotent.*
*(ii) If $G$ is a finite p-group, then $\Phi(G) = G'G^p$, where $G^p$ is the subgroup of $G$ generated by all pth power and $G'$ is the derived subgroup of $G$.*
*(iii) If $G$ is a finite p-group then $G/\Phi(G)$ is a vector space over $\mathbb{Z}_p$*

We are interested in part(iii) of the above theorem. This theorem is the basis of the Burnside basis theorem [11, Theorem 5.40], which guarantees a fixed cardinality for any set of minimal generators for a finite $p$-group.

Now let us think of the MOR cryptosystem as in Section 2.1 where $G$ is a finite $p$-group. Since $\Phi(G)$ is characteristic, hence corresponding to an automorphism $\phi : G \to G$ there is a map $\phi' : G/\Phi(G) \to G/\Phi(G)$. Now $\phi'$ is a linear operator of the vector space $G/\Phi(G)$ over $\mathbb{Z}_p$. As we saw in the earlier section, MOR over linear transformations is only as secure as the El-Gamal cryptosystem over finite fields. So, we conclude that the MOR over finite $p$-groups is also only as secure as the El-Gamal cryptosystem over finite fields.